\documentclass[twocolumn,aps,prb]{revtex4}
 
\usepackage{graphicx}
\usepackage{epsfig}

\begin{document}
\baselineskip11pt
\parindent0pt 
{\bf \sf \huge  Magnetism and Hund's Rule in an Optical Lattice with Cold Fermions} 

\bigskip
 
{\bf \sf K. K\"{a}rkk\"{a}inen,$^{1}$ M. Borgh,$^{1}$  M. Manninen,$^{2}$ \&  \\
S.M. Reimann$^{1}$}

\bigskip

{\sl $^1$Mathematical Physics, LTH, Lund University, \\
SE-22100 Lund, Sweden}

{\sl $^2$NanoScience Center, Department of Physics,\\
FIN-40351 University of Jyv\"{a}skyl\"{a}, Finland}
 
{\bf {\dotfill} 
 
\small

\smallskip

Artificially confined, small quantum systems show a high potential 
for employing quantum physics in technology. 
Ultra-cold atom gases have opened an exciting laboratory in which to explore
many-particle systems that are not accessible in 
conventional atomic or solid state physics\cite{eurphysnews}. 
It appears promising that optical trapping of cold bosonic or fermionic 
atoms will make construction of 
devices with unprecedented precision possible in the future, thereby  
allowing experimenters to make their samples much more ``clean'',
and hence 
more coherent. Trapped atomic quantum gases may thus 
provide an interesting alternative to the quantum dot
nanostructures\cite{reimann02} produced today. 
Optical lattices created by standing laser waves loaded with ultra-cold atoms 
are an example\cite{eurphysnews, jaksch2005} of this. They provide
a unique experimental setup to study {\em artificial crystal structures\/}
with tunable physical parameters. Here we demonstrate that a
two-dimensional optical lattice loaded with 
repulsive, contact-inter\-acting fermions shows a rich and systematic 
magnetic phase diagram.  Trapping a few ($N\le 12$) fermions in each of the 
single-site minima of the optical lattice, we find that  
the {\em shell structure} in these quantum wells determines the magnetism.
In a shallow lattice, the tunneling between the lattice sites is strong, 
and the lattice is non-magnetic. 
For deeper lattices, however, the shell-filling of the single wells 
with fermionic atoms determines the magnetism. 
As a consequence of 
Hund's first rule, the interaction energy is lowered by maximizing 
the number of atoms of the same species. 
This leads to a systematic sequence of non-magnetic, 
ferromagnetic and antiferromagnetic phases.
}

An optical lattice resembles an ``egg box''-like arrangement of single
quantum wells, each confining a small number of atoms. 
Different lattice geometries can be realized (see D.~Jaksch and P.~Zoller\cite{jaksch2005} for a
review).
The tunneling and the localization of atoms in the lattice are
controlled by the lattice depth which can be tuned by changing the 
laser intensity. This allows for a smooth transition from a tightly
bound lattice to a system of nearly free atoms. 
The confined atoms have many internal (hyperfine) states which 
can be manipulated by laser light. 
In a weakly interacting, dilute atom gas, $s$-wave scattering dominates.  
The strength of the short-ranged contact interaction between the atoms   
can be tuned in the vicinity of a Feshbach 
resonance\cite{feshbach1958,inouye1998,courteille1998,roberts1998,duine2004,theis2004}. 

For bosons in an optical lattice, it was possible to realize the 
Mott insulator-superfluid quantum phase 
transition\cite{jaksch98, greiner02,stoferle2004,xu2005}.
More recently, a fermionic many-particle quantum system on a three-dimensional 
lattice was experimentally studied by 
K\"ohl {\it et al.}\cite{kohl2005,kohl2005b}, 
who investigated the transition from a band insulator to a normal state.

The atom dynamics in an optical lattice is often described by the
Hubbard or the Bose-Hubbard models\cite{drummond2005} with the hopping parameter $t$ and
the on-site interaction parameter $U$.  
These models rely on the single-band
approximation which allows at most two atoms in a lattice site in the
fermionic case. 
From an experimental viewpoint, it has been argued that these 
discrete models are ideal for describing contact-interacting atoms 
in an optical lattice\cite{jaksch98}.
In condensed matter physics, the Hubbard model is constantly 
used to describe magnetic correlations in solids. In a deep lattice
(with small $t/U$) for a half-filled band, or one fermion per site, 
the Hubbard model predicts antiferromagnetism. 
In fact, it can be shown that in this case the Hubbard Hamiltonian 
coincides with the antiferromagnetic Heisenberg model of localized 
atoms\cite{vollhardt1994}. 
When the band is filled, each lattice site carries two fermions
with opposite spins, and magnetism can not be observed.
More recently, a SU$(n)$ Hubbard model was applied to describe fermionic atoms 
with $n$ different ``flavors'' or spin states\cite{honerkamp}. 

Here, we go {\it beyond} the lowest-band approximation, applying the 
mean-field approach for a {\it two-component} system:
We consider fermionic atoms with two hyperfine, or spin, species
confined into a two-dimensional (square) lattice, where it is possible to 
trap a few ($N\le 12$) atoms in each single well (at each single site).
We find that depending on the statistics 
and the spin of the trapped atoms, the optical lattices  
can have an intriguing and rich magnetic structure. 
The magnetism follows closely the shell structure in the individual lattice 
wells, emerging as there is an excess of one spin species 
in a single lattice site. The situation is in fact 
analogous to the behavior of quantum dot lattices in semiconductors, 
where the magnetism is governed by the shell structure of the single-dot 
components\cite{mkoskinen2003,pkoskinen2003,kolehmainen}. 

An optical lattice can be created by counter-propagating laser beams. 
The resulting potential is a sinusoidal standing wave
$V_{opt}(\mathbf{r})=V_0(\cos^2(kx)+\cos^2(ky))$, where the amplitude $V_0$ 
is tuned by laser intensity and the wave number $k=2\pi/\lambda$ is
set by the laser wave length\cite{greiner02}. A natural unit for the 
energy is the recoil energy $E_R=\hbar^2k^2/(2m)$, and length is measured 
by the inverse of the wave number. 
(In experiments, an underlying slowly varying 
harmonic confinement is added giving rise to an inhomogeneous filling 
of the lattice and coexistence of insulating and conducting domains. 
This external confinement can be made small and is neglected here.)

We consider a square lattice with two lattice sites in the unit
cell, as sketched in Figure~1(A) and (B). 
One site resides at the center of the unit cell, and the other
one crosses the corner periodically, as indicated in the contour plot of the
optical potential. Note that this is the simplest choice of the unit cell 
allowing for antiferromagnetic alignment of the single-site spins.
The inter-site tunneling can be
tuned by varying the lattice depth $V_0$. With increasing $V_0$ the
atoms become more localized in the lattice sites, the band dispersion
decreases and the shells in the  individual traps are
separated by increasingly large gaps. 
For sufficiently large lattice depth $V_0$, it is the shell structure 
of the the individual atom traps at the lattice sites that determines 
the physical behavior of the lattice. 
\begin{figure}
\includegraphics[width=6.66cm,height=9cm]{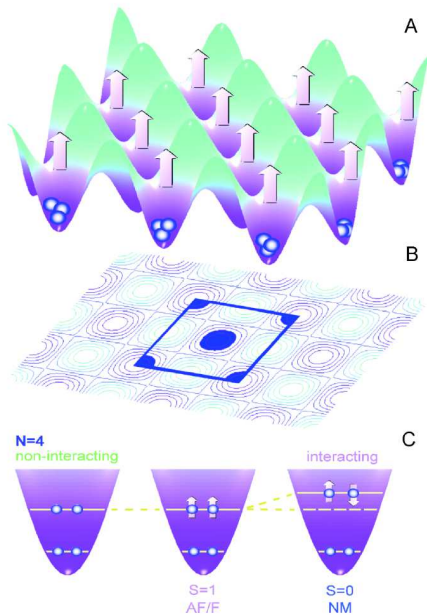}
\caption{(A) Schematic sketch of the optical 
lattice potential, $V_{opt}$, confining a small number of atoms in 
each of the potential minima. (Here, the arrows indicate a ferromagnetic 
ground state as an example). 
(B) shows the contours of $V_{opt}$ and
indicates the unit cell, with one trap at the center of the unit cell, 
and the second one crossing the corner of the cell periodically.
(C) Illustration of Hund's first rule with repulsive contact interactions: 
the energy is lower as the particles become non-interacting
for the same species, i.e. aligned spins (see text).}
\end{figure}

The mechanism leading to magnetic effects in the single quantum wells of 
the lattice is sketched in Figure~1(C). 
In the two-component contact-interacting fermion gas,
due to the Pauli principle there is no mutual interaction 
between atoms of the same species. 
Therefore, the trapped atoms in a degenerate shell can lower their  
interaction energy by maximizing the number of atoms of the same species
-- in other words, by aligning their spins. 
This mechanism in contact-interacting 
atomic systems leads to Hund's first rule\cite{weissbluth1978} and magnetism, 
in close similarity to long-range interacting electronic systems, 
as for example, quantum dots in semiconductor
heterostructures\cite{koskinen1997}. 
Here, however, Hund's rule has a more
dramatic effect 
since it completely removes the interaction between the same
atom species.

A Bose-Einstein condensate of a weakly interacting, dilute gas of 
bosonic atoms is known to be well described by the 
Gross-Pitaevskii equation for the condensate wave function\cite{dalfovo99}. 
Correspondingly, contact-interacting fermions may be approximated by a 
set of Kohn-Sham-like equations with a local effective potential.
We note that in the dilute limit the exchange energy is treated 
exactly\cite{magyar2004}. 
Periodic boundary conditions imply Bloch form for the orbitals, 
$\psi_{n\mathbf{k}\sigma}(\mathbf{r})=\exp(i\mathbf{k}\cdot\mathbf{r})u_{n\mathbf{k}\sigma}(\mathbf{r})$, 
where $n$ labels the band, $\sigma=(\downarrow,\uparrow)$ is the spin index 
and the wave vector $\mathbf{k}$ is confined into the first
Brillouin zone. The periodic functions 
$u_{n\mathbf{k}\sigma}(\mathbf{r})$ satisfy 
\begin{eqnarray}
-\frac{\hbar^2}{2m}(\nabla+i\mathbf{k})^2u_{n\mathbf{k}\sigma}
(\mathbf{r})+[V_{opt}(\mathbf{r})+gn^{\sigma'}(\mathbf{r})]u_{n\mathbf{k}\sigma}(\mathbf{r})
&& \nonumber\\
=\varepsilon_{n\mathbf{k}\sigma}u_{n\mathbf{k}\sigma}(\mathbf{r}),
\quad \sigma \ne \sigma' \quad &&
\label{hf}
\end{eqnarray}
where $m$ is the atom mass, $n^{\sigma}$ is the density of atom species $\sigma$ and $g$ is
the interaction strength. For the latter we have set $g=0.3\ E_R/k^2$ 
in order to stay in the weak-interaction regime, where Eq.~\ref{hf} 
is valid. This was confirmed by  
comparing the mean-field calculations for a single 2D harmonic trap 
with results obtained by exact diagonalization for $N\le 8$  
(see Y.\ Yu {\it et al.,} to be published).
For the Bloch wave vector we use a grid of $5\times 5$ points in the first Brillouin zone, and
the key results were tested with up to $9\times9$ points. 
In the band-structure calculation, the functions $u_{n\mathbf{k}\sigma}(\mathbf{r})$ 
are expanded in a basis of $11\times 11$ plane waves. 
The self-consistent iterations were started with antiferromagnetic and
ferromagnetic initial potentials and 
small random perturbations were added to the initial guesses in order to avoid convergence 
into saddle points of the potential surface. 
\begin{figure}
\includegraphics[width=5.84cm,height=6cm]{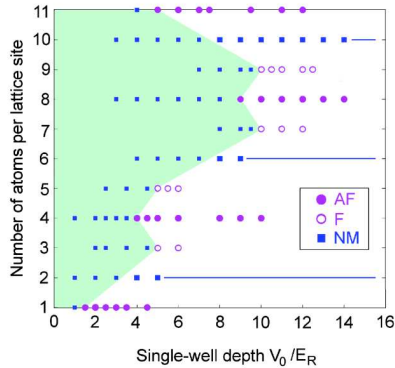}
\caption{Magnetism in a square optical lattice, as a function 
of the lattice depth $V_0$ (determining the tunneling and coupling between the 
single wells) in units of the recoil energy $E_R$, and the particle number (determining the shell
filling). The different 
non-magnetic (NM, {\it blue squares}), ferromagnetic (F, {\it empty circles})
and anti-ferromagnetic (AF, {\it filled circles}) phases are indicated. 
For closed shells (``magic numbers'') at $N=2, 6$ and 10, the lattice is 
non-magnetic. The antiferromagnetic ground state is seen
at mid-shell only, i.e. at $N=1, 4 $ and $8$. 
In between, i.e. at the beginning and end of a shell, the lattice is 
ferromagnetic. The shaded area ({\it green color}) indicates the non-magnetic
region at smaller values of $V_0$.}
\end{figure}

The results of the calculations in the above scheme are displayed 
in Figure~2, which shows the systematics of the magnetism 
as a function of the shell filling of the single  quantum wells in the
lattice as a function of $N$ and $V_0$. The calculated points and the different
ground state configurations are indicated explicitly 
in the figure. Filled and open circles are the antiferromagnetic (AF)
and ferromagnetic (F) states, respectively. 
The non-magnetic region at smaller values of $V_0$ 
in the phase diagram is shaded
and the corresponding states are indicated by squares (NM).
The different magnetic phases follow systematically the filling of the
single-trap shell structure. 

The bottom of the trapping potential 
can be approximated by a harmonic potential with 
$\hbar\omega=\sqrt{4V_0E_R}$. In two dimensions, this leads to the lowest 
closed shells at particle numbers 
$N=$2, 6, and 12 (which are frequently also called ``magic numbers'' 
in the literature, refering back to the conventions in 
nuclear physics)\cite{reimann02}. 
At higher energies, however, the potential has a notch connecting the 
different lattice sites which imposes a square symmetry that breaks 
the three-fold degeneracy of the third, 2$s$1$d$, oscillator shell.
Due to this notch, in our case the closed shells correspond to  
atom numbers $N=$2, 6, {\it 10} and 12. In these cases, the spins in 
the single quantum wells are compensated, and the lattice is non-magnetic. 
However, in between these values, 
at $N=1,4,8$ and 11, the shells are {\it half-filled.} Antiferromagnetic
ordering of spins is observed as a gap is formed at the Fermi-level.
For other atom numbers, the shells are only partially filled, and
therefore the Fermi-levels reside in the middle of a band. These
configurations cannot open a gap. They 
favor ferromagnetism, as the exchange-like effect splits the spin levels.
The lowest band is formed from the 1$s$ levels of the individual
wells. It can be occupied by at most two atoms per
lattice site. By assuming an antiferromagnetic alignment of spins at
half-filled band $N=1$, a gap opens at the Fermi-level since the nearest
similar neighbor is twice as far away as in the ferromagnetic lattice where
every lattice site is identical.
For $N=3,4,5$ and 6 the levels at the 1$p$ shell are occupied. At $N=4$ the 
spin at the 
1$p$ shell is maximized due to Hund's first rule. This mid-shell
configuration favors antiferromagnetism where the energy is reduced by opening
the Fermi-gap. At $N=3$ the shell has only one atom in a $p$-shell 
and at $N=5$ the shell is nearly filled. 
In both these cases the Fermi-energy lies on the band and ferromagnetism is found. 

At the third shell, 2$s$1$d$, a shell closing at $N=10$ with a
non-magnetic phase, and mid-shell fillings  for $N=8$ and $N=11$ 
with antiferromagnetic phases are encountered.
This is attributed to the symmetry-breaking of the $2s1d$ shell that
resides close to the potential junction between different lattice sites.
The $d$-state whose density lobes are not oriented towards the 
nearest neighbors is pushed higher in energy. As a result, the $2s1d$ shell is
split into two sub-shells, one consisting of $2s$ and $1d$ levels and the
other one formed by the higher-lying $1d$ level.
Consequently, we find mid-shell fillings at $N=8$ and 11 leading to antiferromagnetism.

The splitting between the spin bands in the ferromagnetic cases
is of the order of $0.050\ E_R$ which correspond to approximately
$0.048\ k_BT_F$ in a gas of $^{40}$K, where the Fermi-temperature
is\cite{pezze04} 
$T_F=330\ nK$. The typical energy difference between non-magnetic and
ferromagnetic states is $0.001\ E_R$, corresponding to roughly $0.001\ k_BT_F$
in the fermionic potassium gas. These energies correspond to temperatures
which are about a factor of ten lower than the ones achieved 
experimentally. We mention here that an interaction-induced cooling 
mechanism was proposed\cite{werner05} in order to reach the antiferromagnetic 
phase at $N=1$.

\begin{figure}
\includegraphics[width=6.86cm,height=6cm]{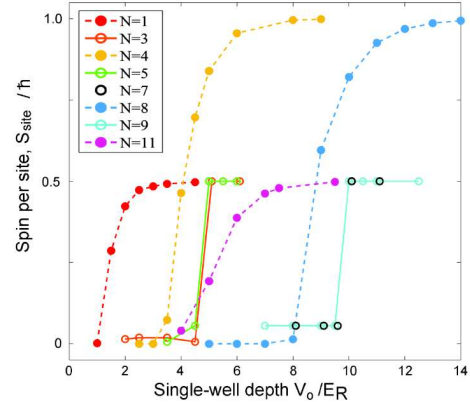}
\caption{Integrated spin density over a single lattice site 
(``spin per site'', $S_{\hbox{\rm site} }/\hbar $), as a function 
of increasing 
lattice depths $V_0$ (in units of the recoil energy $E_R$). 
The different colors correspond to different particle
numbers (shell fillings) of the single traps, as indicated in the inset.}
\end{figure}

The onset of magnetism as a function of $V_0$ can be seen in 
Figure~3 which shows the ''spin per site'', 
obtained by integrating the spin density over a single lattice site,
\begin{equation}
S_{\hbox{\rm site }}=\frac{\hbar}{2}\int_{site}
[n^{\uparrow}(\mathbf{r})-n^{\downarrow}(\mathbf{r})] d\mathbf{r}.
\end{equation}
Generally, the magnetism sets on with increasing $V_0$ as the
number of atoms per lattice site becomes larger and, thus, the spatial
extent of the highest occupied orbital increases. The transition
occurs roughly at the same value of $V_0$ when the Fermi-level resides 
at particular shells. For example, the transition occurs at
$V_0\approx 4.2\ E_R$ for atom numbers $N=3,4$ and 5 as the $1p$ levels are 
occupied. For $N=11$ atoms, antiferromagnetism sets on already 
around $V_0=5\ E_R$. In this case, the highest occupied orbital ($1d$) has only a small
overlap with the corresponding orbital at the neighboring site. 
Therefore, the  
optical lattice has to be shallow before the tunneling between these
levels is large for the magnetism to disappear. 
\begin{figure}
\includegraphics[width=7.14cm,height=6cm]{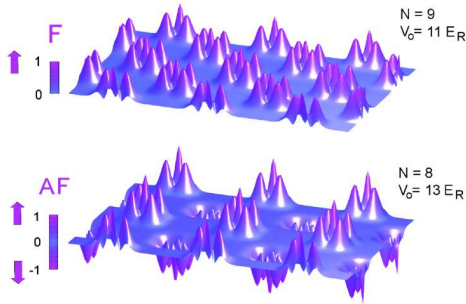}
\caption{Spin density $(n_{\uparrow } - n_{\downarrow })$ 
of the ferromagnetic (F) and antiferromagnetic (AF) states 
for $N=9$ ($V_0=11 E_R$) and $N=8$ ($V_0=13 E_R$), respectively. 
(The amplitude is renormalized to maximum value at unity.)}
\end{figure}

In summary, going beyond the single-band Hubbard model we found that 
two-component cold fermionic atoms in an optical lattice establish a rich 
magnetic phase diagram, and can show analogous effects to the electrons 
in magnetic solids.   
The magnetism in the optical lattice is determined by the number of 
atoms per lattice site, and the depths of the lattice. 
If the lattice is shallow, for strong inter-site tunneling the lattice 
does not show any magnetism. For deeper lattices, where the tunneling is 
small, the total spin of the $N$ atoms at the single sites  
is determined by the shell structure in the lattice minima. 
For closed shells (so-called ``magic numbers'') 
at $N=2,6,10$ and $12$, the single-wells are non-magnetic. 
For contact-interacting repulsive fermions, Hund's first 
rule applies in a particularly 
dramatic way, removing the interaction between the same
atom species.
Antiferromagnetic ordering of the single-site spins was found
when a gap at the Fermi level could open up. This was the case 
at mid-shell, with $N=1, 4, 8,$ or $11$ particles in the single traps. 
Ferromagnetism occurred at the beginning and the end of a shell.

{\bf Acknowledgement:} This work was financially supported by the
Swedish Research Council, the Swedish Foundation for Strategic
Research, the Finnish Academy of Science, and the European Community
project ULTRA-1D (NMP4-CT-2003-505457).
                                                                                
{\bf Correspondence} should be adressed to S.M.~Reimann,
stephanie.reimann@matfys.lth.se


\begin{thebibliography}{99}

\bibitem{eurphysnews} K\"{o}hl, M.\ \& Esslinger, T. Fermionic atoms in an
  optical lattice: a new synthetic material. {\it Europhysics News} {\bf
  37}, 18 (2006).

\bibitem{reimann02} Reimann, S.\ M.\ \& Manninen, M. Electronic
  structure of quantum dots. {\it Rev. Mod. Phys.} {\bf
  74}, 1283 (2002).

\bibitem{jaksch2005} Jaksch, D.\ \& Zoller, P. The cold atom Hubbard toolbox. {\it Annals of Physics} {\bf 315}, 52 (2005).

\bibitem{feshbach1958} Feshbach, H.\ Unified Theory of Nuclear
  Reactions. {\it Annals of Physics} {\bf 5}, 357 (1958).

\bibitem{inouye1998} Inouye, S. Observation of Feshbach resonances in
  a Bose-Einstein condensate. {\it et al.} {\it Nature} {\bf 392}, 151 (1998).

\bibitem{courteille1998} Courteille, Ph.\  {\it et al.} Observation of
  a Feshbach Resonance in Cold Atom Scattering. {\it Phys. Rev. Lett.} {\bf 81},
  69 (1998).

\bibitem{roberts1998} Roberts, J.\ L.\ {\it et al.} Resonant Magnetic
  Field Control of Elastic Scattering in Cold $^{85}$Rb. {\it Phys. Rev. Lett.} {\bf 81},
  5109 (1998).

\bibitem{duine2004} Duine, R.~A. \& Stoof, H.~T.~C. Atom-molecule
  coherence in Bose gases. {\it Physics Reports} {\bf 396}, 115 (2004)

\bibitem{theis2004} Theis, M. Tuning the Scattering Length with an
  Optically Induced Feshbach Resonance. {\it Phys. Rev. Lett.} {\bf 93}, 123001 (2004)

\bibitem{jaksch98} Jaksch, D., Brunder, C., Cirac, J.~I., Gardiner,
  C.~W.\ 
  \& Zoller, P. Cold Bosonic Atoms in Optical Lattices. {\it Phys. Rev. Lett.} {\bf 81}, 3108 (1998).

\bibitem{greiner02} Greiner, M., Mandel, O., Esslinger, T., H\"ansch,
  T.~W.\ 
  \& Bloch, I. Quantum phase transition from a superfluid to a Mott insulator in a gas of ultracold atoms. {\it Nature } {\bf 415}, 39 (2002).

\bibitem{stoferle2004} St\"oferle, T., Moritz, H., Schori, C., K\"ohl,
  M.\ \&
  Esslinger, T. Transition from a Strongly Interacting 1D Superfluid
  to a Mott Insulator. {\it Phys. Rev. Lett.} {\bf 92}, 130403 (2004).

\bibitem{xu2005} Xu, K. Sodium Bose-Einstein condensates in an optical
  lattice. {\it Phys. Rev.} A {\bf 72}, 
043604 (2005).

\bibitem{kohl2005} K\"ohl, M., Moritz, H., St\"oferle, T., G\"unter, K., and
  Esslinger, T. Fermionic Atoms in a Three Dimensional Optical
  Lattice: Observing Fermi Surfaces, Dynamics, and Interactions. {\it Phys. Rev. Lett.} {\bf 94}, 080403 (2005)

\bibitem{kohl2005b} K\"ohl, M., G\"unter, K., St\"oferle, T., Moritz,
  H.\ \&
  Esslinger, T. Strongly Interacting Atoms and Molecules in a 3D
  Optical Lattice. http://arxiv.org/abs/cond-mat/0605099 (2006)

\bibitem{drummond2005} Drummond, P.~D., Corney, J.~F., Liu, X.-J. \& 
  Hu, H. Ultra-cold fermions in optical lattices. 
{\it Journal of Modern Optics} {\bf 52} No. 16, 2261-2268 (2005). 

\bibitem{vollhardt1994} Vollhardt, D.\ in Broglia, R.~A., Schrieffer,
  J.~R.\ \& Bortignon, P.~F. (editors) {\it Perspectives in Many-Body
  Physics.} North-Holland, Amsterdam, (1994)

\bibitem{honerkamp} Honerkamp, C., \& Hofstetter, W., Ultracold
  Fermions and the SU($N$) Hubbard Model. {\it Phys. Rev. Lett.} {\bf
  92}, 170403 (2004).

\bibitem{mkoskinen2003} Koskinen, M., Reimann, S.~M., \& Manninen
  M. Spontaneous Magnetism of Quantum Dot Lattices. 
  {\it Phys. Rev. Lett.} {\bf 90}, 066802 (2003). 

\bibitem{pkoskinen2003} Koskinen, P., Sapienza, L.\ \& Manninen,
  M. Tight-Binding Model for Spontaneous Magnetism of Quantum Dot
  Lattices. {\it Physica Scripta}  {\bf 68}, 74 (2003).

\bibitem{kolehmainen} Kolehmainen, J., Reimann, S.~M., Koskinen, M.\ \&
  Manninen, M. Magnetic interaction between coupled quantum dots {\it
  Eur.\ Phys.\ J.\ B} {\bf 13}, 731 (2000).

\bibitem{weissbluth1978} Weissbluth, M. {\it Atoms and Molecules.}
  Academic Press, N.Y., (1978).

\bibitem{koskinen1997} Koskinen, M., Manninen, M.\ \& Reimann,
  S.~M. Hund's Rules and Spin Density Waves in Quantum Dots. {\it
    Phys. Rev. Lett.} {\bf 79}, 1389 (1997).

\bibitem{dalfovo99} Dalfovo, F., Giorgini, S., Pitaevskii, L.~P.\ \&
  Stringari, S. Theory of Bose-Einstein condensation in trapped
  gases. {\it Rev. Mod. Phys.} {\bf 71}, 463 (1999).

\bibitem{magyar2004} Magyar, R.~J. \& Burke, K. Density-functional
  theory in one dimension for contact-interacting fermions. {\it Phys.\
    Rev.\ A} {\bf 70}, 032508 (2004).

\bibitem{pezze04}  Pezz\`e, L.\ {\it et al.} Insulating Behavior of a
  Trapped Ideal Fermi Gas. {\it Phys. Rev. Lett.} {\bf 93}, 120401 (2004).

\bibitem{werner05} Werner, F. , Parcollet, O., Georges, A. \& 
  Hassan, S.~R. Interaction-Induced Adiabatic Cooling and
  Antiferromagnetism of Cold Fermions in Optical Lattices. {\it
  Phys. Rev. Lett.} {\bf 95}, 056401 (2005).

\end{thebibliography}
\end{document}